\documentclass[twocolumn,pre,showpacs]{revtex4}
\usepackage{amssymb}
\usepackage{amsmath}
\usepackage{graphicx}

\input{tcilatex}

\begin{document}

\title{Quantum-Classical Transition of Photon-Carnot Engine \\
Induced by Quantum Decoherence }
\author{H.T. Quan, P. Zhang and C.P. Sun }
\email{suncp@itp.ac.cn}
\homepage{http://www.itp.ac.cn/~suncp}
\affiliation{Institute of Theoretical Physics, Chinese Academy of Sciences, Beijing,
100080, China}

\begin{abstract}
We study the physical implementation of the Photon Carnot engine (PCE) based
on the cavity QED system [M. Scully et al, Science, \textbf{299}, 862
(2003)]. Here, we analyze two decoherence mechanisms for the more practical
systems of PCE, the dissipation of photon field and the pure dephasing of
the input atoms. As a result we find that (I) the PCE can work well to some
extent even in the existence of the cavity loss (photon dissipation); and
(II) the short-time atomic dephasing, which can destroy the PCE, is a fatal
problem to be overcome.
\end{abstract}

\maketitle

\section{INTRODUCTION}

Recently many investigations have been carried out to explore various
possibilities of building Carnot (or Otto, etc.) heat engines in some
\textquotedblleft quantum way". It is expected that, by taking the
advantages of quantum coherence, such quantum heat engines (QHE) using
quantum matter as working substance can improve work extraction as well as
the working efficiency in a thermodynamic cycle \cite%
{Scully1,Scully3,Scully4,Opatrny}. Marlan O. Scully and his collaborators
proposed and studied a QHE based on a cavity QED system \cite{Scully-book,
Scully-science, Zubairy-book}, namely, a photon-Carnot engine (PCE) \cite%
{Lee}. The working substance of their PCE is a lot of single-mode photons
radiated from the partially coherent atoms. In their model, the walls of the
cavity are assumed to be ideal, i.e., the cavity loss are disregarded.

In practice, however, the walls can not perfectly reflect the photons (the
cavity loss are not negligible), and the atomic dephasing of the input atoms
are inevitable due to its coupling to the environment when passing through
the cavity. To focus on the essence of the problem we only
phenomenologically consider the pure dephasing effect \cite{Decoherence
mechanism} in this paper. A question then follows naturally: How does the
photon dissipation and atomic dephasing influence the efficiency of the PCE?
In this paper, by analyzing a more realistic cavity QED system, we revised
the PCE model proposed by Scully and his collaborators. We find the
efficiency of the PCE decrease when the cavity quality $Q$ becomes smaller
(cavity loss becomes more strong); when the atomic dephasing happens, though
the atomic energy conserves, the quantum features of the PCE are demolished
and then the QHE becomes a classical one.

Our investigation is significant in two aspects. On the one hand, our
results confirm the robustness of the PCE proposed in Refs. \cite%
{Scully-book,Scully-science,Zubairy-book}, which can still work well even in
the existence of not too strong cavity loss. On the other hand, our results
demonstrate the quantum-classical transition of the PCE due to quantum
dephasing, which agrees well with our intuition, the efficiency of the PCE
decrease due to atomic dephasing. These predictions can not only help us
better the understanding of the basic concepts of thermodynamics,
statistical mechanics, but also help us to optimize the system parameters in
future experiments of PCE. It is also of interest that the efficiency of the
PCE in a Carnot cycle can be used to measure the quantum coherence of the
input atoms and characterize the quantum-classical transition of the PCE. 
%Figure 1
%\begin{figure}[h]
\begin{figure}[tbp]
\begin{center}
\includegraphics[bb=50 90 554 755, width=5cm, clip]{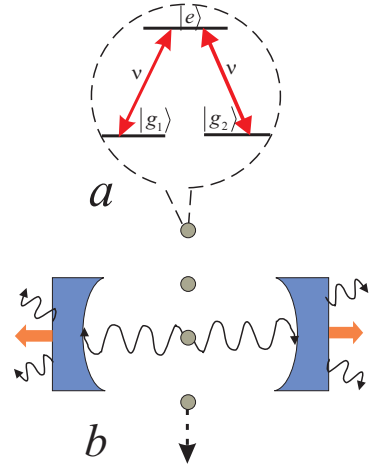}
\end{center}
\caption{(color online) The cavity QED model of our QHE. Three-level atoms,
with quantum coherence in the two degenerate ground states $\left\vert
g_{1}\right\rangle $ and $\left\vert g_{2}\right\rangle $, flow through the
cavity and interact with the resonant photon field in the cavity.}
\end{figure}

\section{CAVITY QED MODEL OF QHE REVISED}

The PCE we consider here is similar to that proposed in Refs. \cite%
{Scully-book,Scully-science,Zubairy-book} (see the schematic illustration in
Fig. 1). In our PCE model, the ground states $\left\vert g_{1}\right\rangle $%
\ and $\left\vert g_{2}\right\rangle $ are accurately degenerate. The
atom-photon coupling is described by the Hamiltonian \cite{Scully2, Orszag}%
\begin{equation}
H_{I}=\hbar \lambda \left\vert e\right\rangle (\frac{\left\langle
g_{1}\right\vert +\left\langle g_{2}\right\vert }{\sqrt{2}})a+h.c..
\end{equation}%
Here, $\nu $ is the level spacing between the excited state $\left\vert
e\right\rangle $ and the ground states; $a$ is the annihilation operator of
the resonant photon field and $\lambda $ is the atom-field coupling constant.

If there were no photon dissipation and atomic dephasing, $H_{I}$ would
completely govern the pure quantum evolution. Let $\left\vert m\right\rangle 
$, $m=0,1,2,\cdots $ denote the Fock state of the photon field. In every
invariant subspace $V_{m}$, which is spanned by the ordered basis vectors $%
\{\left\vert e\right\rangle \otimes \left\vert m-1\right\rangle ,\left\vert
g_{1}\right\rangle \otimes \left\vert m\right\rangle ,\left\vert
g_{2}\right\rangle \otimes \left\vert m\right\rangle \}$, the evolution
operator $U\left( \tau \right) =\exp (-iH_{I}t)$ can be expressed as a
quasi-diagonal matrix with the diagonal blocks%
\begin{equation}
U_{m}\left( \tau \right) =\left[ 
\begin{array}{ccc}
C_{m}(\tau ) & \frac{-i}{\sqrt{2}}S_{m}(\tau ) & -\frac{i}{\sqrt{2}}%
S_{m}(\tau ) \\ 
\frac{-i}{\sqrt{2}}S_{m}(\tau ) & C_{m}^{2}(\frac{\tau }{2}) & -S_{m}^{2}(%
\frac{\tau }{2}) \\ 
\frac{-i}{\sqrt{2}}S_{m}(\tau ) & -S_{m}^{2}(\frac{\tau }{2}) & C_{m}^{2}(%
\frac{\tau }{2})%
\end{array}%
\right] .  \label{1}
\end{equation}%
Here, $C_{m}(\tau )=\cos \left( \lambda \sqrt{m}\tau \right) $ and $%
S_{m}(\tau )=\sin \left( \lambda \sqrt{m}\tau \right) $. In obtaining above
explicit expressions for the matrix elements of $U_{m}\left( \tau \right) $,
we have used the following technique. By writing%
\begin{equation*}
\left\vert G\right\rangle =\frac{\left\langle g_{1}\right\vert +\left\langle
g_{2}\right\vert }{\sqrt{2}},
\end{equation*}%
we find%
\begin{equation}
H_{I}=\hbar \lambda \left\vert e\right\rangle \left\langle G\right\vert
a+h.c.
\end{equation}%
is, in fact, the Hamiltonian for the Jaynes-Cummings model at resonance, and
then we can exactly solve the problem of time evolution in each subspace $%
V_{m}$ following the method in Refs. \cite{Scully2, Orszag}.

We denote the initial density matrix of the photon field and the atom
ensemble as $\rho _{L}\left( t_{i}\right) $ and $\rho _{A}\left(
t_{i}\right) $. The reduced density matrix $\rho _{L}\left( t\right) $ of
the radiation field will evolve according to the \textquotedblleft
super-operator" $M\left( \tau \right) $ defined by%
\begin{equation}
M\left( \tau \right) \rho _{L}\left( t_{i}\right) =Tr_{A}[U\left( \tau
\right) \rho _{L}\left( t_{i}\right) \otimes \rho _{A}\left( t_{i}\right)
U^{\dagger }\left( \tau \right) ],
\end{equation}%
where $Tr_{A}$ means tracing over the atomic variable. The atoms pass
through the cavity at the rate $r$. Then we can write the known master
equation at zero temperature \cite{Orszag,Scully2} as 
\begin{equation}
\frac{d}{dt}\rho _{L}\left( t\right) \approx r[M\left( \tau \right) -1]\rho
_{L}\left( t\right) +L\rho _{L}\left( t\right) .  \label{2}
\end{equation}%
Here, we have made the approximation $\ln \left[ M\left( \tau \right) \right]
\approx M\left( \tau \right) -1$ for a short time $\tau $ (the reliability
of this approximation can be seen from later parameter estimation $\tau \sim
10^{-10}$ s), and the cavity loss term is defined as%
\begin{equation}
L\rho _{L}\left( t\right) =(\frac{\nu }{2Q})[2a\rho _{L}(t)a^{\dagger
}-(a^{\dagger }a\rho _{L}(t)+h.c.)]
\end{equation}%
with the cavity quality factor $Q$.

Inspired by the concept of \textquotedblleft phaseonium" in Ref. \cite%
{Scully-science}, we prepare the atoms initially in a partially coherent
state%
\begin{equation}
\rho _{A}\left( 0\right) =p_{e}\left\vert e\right\rangle \left\langle
e\right\vert +\left\vert g\right\rangle \left\langle g\right\vert .
\end{equation}%
Here, $\left\vert g\right\rangle \left\langle g\right\vert $ contains the
superposition of the ground states%
\begin{eqnarray}
\left\vert g\right\rangle \left\langle g\right\vert &=&\left\vert
c_{1}\right\vert ^{2}\left\vert g_{1}\right\rangle \left\langle
g_{1}\right\vert +\left\vert c_{2}\right\vert ^{2}\left\vert
g_{2}\right\rangle \left\langle g_{2}\right\vert \\
&&+c_{1}c_{2}^{\ast }\left\vert g_{1}\right\rangle \left\langle
g_{2}\right\vert +c_{1}^{\ast }c_{2}\left\vert g_{2}\right\rangle
\left\langle g_{1}\right\vert ;  \notag
\end{eqnarray}%
$p_{e}$ is the probability distribution in the excited state $\left\vert
e\right\rangle $. $\rho _{A}\left( 0\right) $ has been normalized to unit,
i.e.,%
\begin{equation}
p_{e}+\left\vert c_{1}\right\vert ^{2}+\left\vert c_{2}\right\vert ^{2}=1.
\end{equation}%
It is well known that because the atomic dephasing is much more rapid \cite%
{Orszag, Decoherence time, Dephasing time} than the quantum dissipation of
the atoms, quantum dephasing is the dominating decoherence mechanism in
short time. Hence, we only phenomenologically consider the atomic dephasing
when the atoms pass through the cavity. We replace the pure state $%
\left\vert g\right\rangle \left\langle g\right\vert $ in $\rho _{A}\left(
0\right) $\ with the mixture%
\begin{equation}
\rho _{g}=\sum_{k=1,2}\left\vert c_{k}\right\vert ^{2}\left\vert
g_{k}\right\rangle \left\langle g_{k}\right\vert +\xi c_{1}c_{2}^{\ast
}\left\vert g_{1}\right\rangle \left\langle g_{2}\right\vert +h.c.,
\label{4}
\end{equation}%
where $\xi $ is the so-called decoherence factor satisfying $|\xi |\leq 1$ 
\cite{Decoherence mechanism}. The complete coherence is characterized by $%
\xi =1$, while the complete decoherence is by $\xi =0$. With above
considerations, we depict the atomic dephasing by changing the initial state
from $\rho _{A}\left( 0\right) $ to%
\begin{equation}
\rho _{D}=p_{e}\left\vert e\right\rangle \left\langle e\right\vert +\rho
_{g}.
\end{equation}

\section{PHOTON-\textquotedblleft WORKING FLUID" AT STATIONARY STATE
DESCRIBED BY AN EFFECTIVE TEMPERATURE}

Next we consider a similar cycle as that in Ref. \cite{Scully-book}. During
the isothermal expansion process the input atoms, prepared in a state
slightly deviating from the thermal equilibrium with some quantum coherence,
serves as a high temperature energy source, while during the isothermal
compression process the input atoms, prepared in thermal equilibrium state,
serves as the entropy sink.\ For a short interaction time $\tau $, we have
(the reliability of this approximation can also be seen from later
parameters estimation $\sqrt{m}\lambda \tau \sim 10^{-1}$)%
\begin{eqnarray}
C_{m}(\tau ) &\simeq &1-\frac{1}{2}m\lambda ^{2}\tau ^{2}, \\
S_{m}(\tau ) &\simeq &\sqrt{m}\lambda \tau .  \notag
\end{eqnarray}%
Then we obtain the equation of motion for the average photon number $%
\left\langle n\left( t\right) \right\rangle $%
\begin{equation}
\frac{d}{dt}\left\langle n\left( t\right) \right\rangle =\mu \lbrack
(2p_{e}-\theta )\left\langle n\left( t\right) \right\rangle +2p_{e}]-\frac{%
\nu }{Q}\left\langle n\left( t\right) \right\rangle ,  \label{5}
\end{equation}%
where%
\begin{eqnarray}
\mu &=&\frac{r\lambda ^{2}\tau ^{2}}{2}, \\
\theta &=&|c_{1}|^{2}+|c_{2}|^{2}+2Re(\xi c_{1}c_{2}^{\ast }),  \notag
\end{eqnarray}%
and Re$(\xi c_{1}c_{2}^{\ast })$ is the real part of $\xi c_{1}c_{2}^{\ast }$%
.

In the thermal equilibrium state, the atomic probability distributions $%
p_{e} $, $|c_{1}|^{2}$ and $|c_{2}|^{2}$\ satisfy%
\begin{equation}
\frac{p_{e}}{|c_{1}|^{2}}=\frac{p_{e}}{|c_{2}|^{2}}=\exp (-\frac{\hbar \nu }{%
kT}),
\end{equation}%
where $k$ is the Boltzmann constant and $T$ is the temperature of the
thermalized atoms. Since the relaxation time of the radiation field is very
short, in the following analysis, we will use the equilibrium state solution%
\begin{equation}
\left\langle n_{E}\right\rangle =\frac{n}{1+\zeta (T)}
\end{equation}%
of Eq. (\ref{5}) to replace the time-dependent average photon number $%
\left\langle n\left( t\right) \right\rangle $. Here, $n$\ is the average
photon number%
\begin{equation}
n=\frac{2p_{e}}{|c_{1}|^{2}+|c_{2}|^{2}-2p_{e}}
\end{equation}%
in the absence of atomic coherence and cavity loss, and%
\begin{equation}
\zeta (T)=\frac{n}{p_{e}}[\text{Re}(\xi c_{1}c_{2}^{\ast })+\frac{\nu }{2\mu
Q}]  \label{7}
\end{equation}%
is a temperature dependent parameter concerning the cavity loss as well as
the atomic dephasing.

We imagine the radiation field also obeys a virtual Bose distribution%
\begin{equation}
\left\langle n_{E}\right\rangle =\frac{1}{\exp [\hbar \nu /(kT^{\prime })]-1}
\end{equation}%
with an effective temperature $T^{\prime }$. In high temperature limit, we
approximately have (for the microwave cavity QED system and circuit QED,
this approximation is reliable when the effective temperature $T^{\prime }$
is at room temperature or higher)%
\begin{equation}
\left\langle n_{E}\right\rangle \approx \frac{kT^{\prime }}{\hbar \nu },%
\text{ }n\approx \frac{kT}{\hbar \nu }.  \label{7.5}
\end{equation}%
Therefore $T^{\prime }$ can be approximately determined as \cite%
{Scully-science}%
\begin{equation}
T^{\prime }=\frac{T}{1+\zeta (T)}.  \label{8}
\end{equation}%
It can be seen that the effective temperature $T^{\prime }$ being different
from $T$\ is due to the atomic coherence as well as the cavity loss.
Obviously, when $Q\rightarrow \infty $ and $\text{Re}(\xi c_{1}c_{2}^{\ast
})=0$, the effective temperature $T^{\prime }$ approaches $T$. I.e., the
effective temperature becomes equal to the temperature of the input atoms
when the atomic coherence cancels and the cavity is perfect. 
\begin{figure}[h]
\begin{center}
\includegraphics[bb=23 359 587 740, width=6 cm, clip]{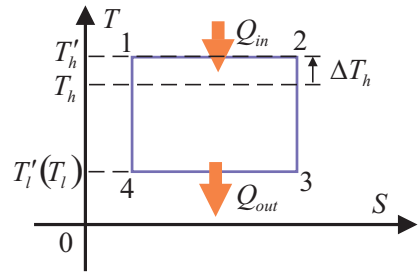}
\end{center}
\caption{(color online) Temperature-entropy diagram for the photon-Carnot
cycle. Here we consider neither the cavity loss nor the atomic dephasing.
i.e., $\protect\xi =1$, $Q\rightarrow \infty $. In addition, we consider the
special case of the phase angle Arg$\left( c_{1}c_{2}^{\ast }\right) =%
\protect\pi $. As a result the effective temperature $T_{h}^{\prime }$ is
higher than $T_{h}$ by the amount $\Delta T_{h}=n_{h}T_{h}\left\vert
c_{1}^{h}c_{2}^{h\ast }\right\vert /P_{e}^{h}$.}
\end{figure}

\section{THERMODYNAMIC CYCLE OF THE QHE WITH QUANTUM MATTER AS THE WORKING
SUBSTANCE}

The Carnot cycle of our QHE consists of two isothermal and two adiabatic
processes \cite{Scully-book} (see Fig. 2). We use the subscripts (or
superscript) $h$ and $l$ to indicate the isothermal expansion and the\
isothermal compression processes hereafter. E.g., $n_{h}$ and $n_{l}$
represent $n$ in the isothermal expansion and the isothermal compression
processes respectively, i.e.,%
\begin{equation}
n_{i}=\frac{2p_{e}^{i}}{|c_{1}^{i}|^{2}+|c_{2}^{i}|^{2}-2p_{e}^{i}},i=h,l.
\end{equation}%
During the isothermal expansion process, from a thermal state 1\textbf{\ }of
the photon field to another 2, the input three-level atoms are prepared with
quantum coherence of the ground states $\rho _{A}^{h}\left( 0\right) \equiv
\rho _{A}\left( 0\right) $. But during the isothermal compression process
from 3 to 4, the input atoms are prepared in a thermalized state, i.e.,%
\begin{equation}
\rho _{A}^{l}\left( 0\right) =p_{e}^{l}\left\vert e\right\rangle
\left\langle e\right\vert +|c_{1}^{l}|^{2}\left\vert g_{1}\right\rangle
\left\langle g_{1}\right\vert +|c_{2}^{l}|^{2}\left\vert g_{2}\right\rangle
\left\langle g_{2}\right\vert .
\end{equation}%
>From above facts and Eq. (\ref{7}) we know, in the isothermal expansion
process from 1 to 2%
\begin{equation}
\zeta _{h}(T_{h})=\frac{n_{h}}{p_{e}^{h}}[Re(\xi c_{1}^{h}c_{2}^{h\ast })+%
\frac{\nu _{h}}{2\mu Q}],  \label{10}
\end{equation}%
but in the isothermal compression process%
\begin{equation}
\zeta _{l}(T_{l})=\frac{\nu _{l}n_{l}}{2\mu Qp_{e}^{l}}.  \label{11}
\end{equation}

We apply the entropy expression%
\begin{equation}
S_{i}=k\ln \left( \left\langle n_{E}^{i}\right\rangle +1\right) +\frac{\hbar
\nu _{i}\left\langle n_{E}^{i}\right\rangle }{T_{i}^{\prime }}\text{\ \ }%
\left( i=h,l\right)  \label{9}
\end{equation}%
of the radiation field to calculate the heat transfer. In a Carnot circle,
the work done by the radiation field is $\Delta W=Q_{in}-Q_{out}$. Here%
\begin{equation}
Q_{in}=T_{h}^{\prime }[S_{h}\left( 2\right) -S_{h}\left( 1\right) ],
\end{equation}%
and%
\begin{equation}
Q_{out}=T_{l}^{\prime }[S_{l}\left( 3\right) -S_{l}\left( 4\right) ]
\end{equation}%
are respectively the heat absorbed into the cavity during the isothermal
expansion process from 1 to 2 and the heat released out of the cavity during
the isothermal compression process from 3 to 4.

Similar to the cycle in Ref. \cite{Scully-book}, the frequency of the
radiation field, i.e., the resonant mode of the cavity, is assumed to change
slightly form 1 to 2, i.e.,%
\begin{equation}
\frac{\left[ \nu \left( 1\right) -\nu \left( 2\right) \right] }{\nu \left(
1\right) }\ll 1.
\end{equation}%
Namely, we can make the approximations%
\begin{eqnarray}
\nu \left( 1\right)  &\approx &\nu _{l}\approx \nu \left( 2\right) , \\
\nu \left( 3\right)  &\approx &\nu _{h}\approx \nu \left( 4\right) .  \notag
\end{eqnarray}%
In the adiabatic process, the average photon number does not change, i.e.,%
\begin{eqnarray}
\left\langle n_{E}^{_{h}}\left( 2\right) \right\rangle  &=&\left\langle
n_{E}^{_{l}}\left( 3\right) \right\rangle , \\
\left\langle n_{E}^{_{h}}\left( 4\right) \right\rangle  &=&\left\langle
n_{E}^{_{l}}\left( 1\right) \right\rangle .  \notag
\end{eqnarray}%
Form Eq. (\ref{7.5}), it follows that%
\begin{equation}
\frac{\nu \left( 1\right) }{T_{h}^{\prime }}=\frac{\nu \left( 4\right) }{%
T_{l}^{\prime }},\text{ }\frac{\nu \left( 2\right) }{T_{h}^{\prime }}=\frac{%
\nu \left( 3\right) }{T_{l}^{\prime }}.
\end{equation}%
From these observations and Eq. (\ref{9}) we find that%
\begin{equation*}
S_{h}\left( 2\right) -S_{h}\left( 1\right) =S_{l}\left( 3\right)
-S_{l}\left( 4\right) .
\end{equation*}%
Therefore, in the high temperature limit, the PCE efficiency $\eta =\left(
Q_{in}-Q_{out}\right) /Q_{in}$ can be explicitly expressed as $\eta
=1-T_{l}^{\prime }/T_{h}^{\prime }$ \cite{Scully-book} or%
\begin{equation}
\eta =1-\left[ \frac{1+\zeta _{h}(T_{h})}{1+\zeta _{l}(T_{l})}\right] \frac{%
T_{l}}{T_{h}}.  \label{12}
\end{equation}%
Based on above results we are now able to analyze the effects of the two
decoherence mechanisms separately.

Firstly, to focus on the cavity loss (photon dissipation), we consider the
ideal case with no atomic dephasing, $\xi =1$. In the ideal case the cavity
loss is negligible, i.e.,$Q\rightarrow \infty $, the efficiency (\ref{12})
is reduced to%
\begin{equation}
\eta =1-[1+\frac{n_{h}}{p_{e}^{h}}Re(c_{1}^{h}c_{2}^{h\ast })]\frac{T_{l}}{%
T_{h}},
\end{equation}%
which agrees with the result in Ref. \cite{Scully-science}. It seems that,
in principle, the PCE can extract positive work from a single heat bath if
we control the phase angle $\theta =$Arg$(c_{1}^{h}c_{2}^{h\ast })$
properly, e.g., $\theta =\pi $. This shows the advantage of the
\textquotedblleft quantum fuel"--we can extract more work from the
\textquotedblleft quantum fuel" than from the classical fuel. However, in
the case the cavity is \textquotedblleft extreme bad", i.e., the cavity
quality factor is vanishingly small $Q\rightarrow 0$, the efficiency
decrease to zero (this can be verified easily)%
\begin{equation}
\eta \rightarrow 1-\frac{\nu _{h}n_{h}p_{e}^{l}}{\nu _{l}n_{l}p_{e}^{h}}%
\frac{T_{l}}{T_{h}}\approx 0.  \label{xxx}
\end{equation}%
Namely, the PCE is destroyed by the strong loss of the cavity. This can also
be understood intuitively from the Eq. (\ref{8}). When the quality factor of
the cavity becomes so small that few photons can stay stably in the cavity.
Accordingly, both the two effective temperatures $T_{h}^{\prime }$ and $%
T_{l}^{\prime }$ becomes vanishingly small, and thus no work can be done by
the \textquotedblleft working substance". As a result, the efficiency of the
PCE decrease to zero. Mathematically, it can be verified that $\eta $ is a
monotonically increasing function of $Q$, and the efficiency of the PCE
decrease to zero when $Q$ becomes vanishingly small.

Secondly we consider the pure atomic dephasing. From Refs. \cite{Orszag,
Decoherence time, Dephasing time} we know that the dephasing time is much
shorter than the atom and cavity lifetimes.\ After the atom interacting with
the environment for a short time $\tau $, the term Re$(\xi
c_{1}^{h}c_{2}^{h\ast })$ concerning atomic coherence becomes vanishingly
small. We can properly assume $\xi $ decrease to zero in the short time $%
\tau $. Then the efficiency (\ref{12}) of the PCE becomes%
\begin{equation}
\eta =1-\left[ \frac{1+\nu _{h}n_{h}/(2\mu p_{e}^{h}Q)}{1+\nu
_{l}n_{l}/(2\mu p_{e}^{l}Q)}\right] \frac{T_{l}}{T_{h}}.  \label{14}
\end{equation}%
In principle, if the cavity is ideal, $Q\rightarrow \infty $, we regain the
maximum classical Carnot efficiency\ $\eta =1-T_{l}/T_{h}$ from the
efficiency (\ref{14}) of the PCE. It turns out that, without cavity loss,
the complete dephasing of the atoms makes the PCE become an ideal
(reversible)\ classical heat engine. This demonstrate the quantum-classical
transition due to the atomic dephasing. Similarly, in the case of
\textquotedblleft extremely bad" cavity, $Q\rightarrow 0$, the efficiency (%
\ref{14}) decrease to zero (\ref{xxx}).

%Figure 3
\begin{figure}[h]
\begin{center}
\includegraphics[bb=31 358 572 769, width=6 cm, clip]{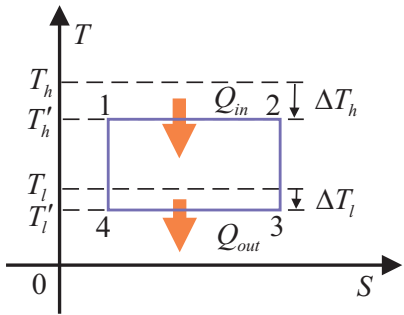}
\end{center}
\caption{(color online) Temperature-entropy diagram for the photon-Carnot
cycle. The atomic coherence are demolished due to the atomic dephasing,
i.e., $\protect\xi =0$. The cavity loss makes the effective cavity
temperature decrease. Here, $\Delta T_{h}=n_{h}T_{h}\protect\nu _{h}/\left( 2%
\protect\mu P_{e}^{l}Q\right) $, and $\Delta T_{l}=n_{l}T_{l}\protect\nu %
_{l}/\left( 2\protect\mu P_{e}^{h}Q\right) $.}
\end{figure}
\ Finally we analyze the positive-work condition of this PCE \cite{ours,
Kieu, Kieu1}, under which positive-work can be extracted. From Eq. (\ref{12}%
) we know that the positive work condition $\Delta W=Q_{in}-Q_{out}>0$ of
this QHE is 
\begin{equation}
T_{h}>\frac{1+\zeta _{h}(T_{h})}{1+\zeta _{l}(T_{l})}T_{l},  \label{15}
\end{equation}%
where $[1+\zeta _{h}(T_{h})]/[1+\zeta _{l}(T_{l})]$ can be either less or
greater than unit. It is counterintuitive when $[1+\zeta
_{h}(T_{h})]/[1+\zeta _{l}(T_{l})]$\ is smaller than unit, and the same
novel result occurs in Ref. \cite{Scully-science}. These novel results
originate from the fact that the atoms are not prepared in the thermal
equilibrium state in the isothermal expansion process. In other words, the
initial state with partial quantum coherence is out of thermal equilibrium,
and thus the \textquotedblleft temperature" $T_{h}$\ of the input atoms in
the isothermal expansion process should not be regarded as a real
thermodynamic temperature essentially. We also would like to emphasize that
the second law of thermodynamics is not violated when%
\begin{equation*}
\frac{1+\zeta _{h}(T_{h})}{1+\zeta _{l}(T_{l})}<1,
\end{equation*}%
for it would take energy from an external source to prepare the atomic
coherence \cite{Zubairy-book}. In an overall consideration, the extra energy
cost for preparing the atomic coherence would prevent the second law from
being violated.

\section{EXPERIMENTAL FEASIBILITY}

Finally, we would like to estimate the efficiency $\eta $ according to a set
of experimentally accessible parameters. For different material and cavity
parameters \cite{Blais} the discussions are listed below.

Firstly, for the optical cavity QED system \cite{Blais}, the resonance
frequency $\nu _{h}\left( \nu _{l}\right) \sim 10^{14}$ Hz, the atom-field
coupling constant $\lambda \sim 10^{8}$ Hz and the quality factor is less
than $10^{8}$. We take a set of reasonable values \cite{Scully-science,
Scully-book}: $n_{h}\left( n_{l}\right) \sim 10^{2}$, $\lambda \tau \sqrt{n}%
\sim 10^{-1}$. Thus we get order of magnitude of the interaction time $\tau
\sim 10^{-10}$ s. To require only one atom in the cavity once, we have $%
r\leq \left( 1/\tau \right) \sim 10^{10}$ /s. Based on above estimation, we
get $\mu =r\lambda ^{2}\tau ^{2}/2\leq 10^{6}\ $/s. Hence, the term $\nu
_{h}/\left( 2\mu Q\right) $\ in Eq. (\ref{10}) has the order of magnitude
greater than $10^{0}$, but the order of magnitude of the other part Re$(\xi
c_{1}^{h}c_{2}^{h\ast })$\ in Eq. (\ref{10}) is less than $10^{-1}$\ due to
the atomic dephasing $\xi \leq 1$. It can be seen, for the optical QED
system, when considering the practical experimental parameters, we can well
approximately disregard the term Re$(\xi c_{1}^{h}c_{2}^{h\ast })$\
concerning quantum coherence in calculating the efficiency.

Secondly, for the microwave cavity QED system \cite{Blais}, the resonance
frequency $\nu _{h}\left( \nu _{l}\right) \sim 10^{10}$ Hz, the atom-field
coupling constant $\lambda \sim 10^{4}$ Hz and the quality factor is less
than $10^{9}$. After similar analysis, we find the term $\nu _{h}/\left(
2\mu Q\right) $\ in Eq. (\ref{10}) has the order of magnitude greater than $%
10^{-1}$, but the order of the other part Re$(\xi c_{1}^{h}c_{2}^{h\ast })$\
in Eq. (\ref{10}) is less than $10^{-1}$\ due to the atomic dephasing $\xi
\leq 1$. Thus the conclusion for optical cavity system remains valid for
microwave cavity system.

Thirdly, we consider the circuit QED based on superconducting Josephson
junction systems \cite{circuit, Blais}. Here, the coupling between the
charge qubit [a Cooper pair box (CPB)] and a quantum transmission line is in
the similar way as the photon-atom coupling in cavity QED system. The
passage of atoms through the cavity can be simulated by the periodic switch
-off and -on of the on chip-interaction \cite{Peng,Zoller}. The experimental
parameters can also be found in Ref. \cite{Blais}: the resonance frequency $%
\nu _{h}\left( \nu _{l}\right) \sim 10^{10}$ Hz; the atom-field coupling
constant $\lambda \sim 10^{8}$ Hz and the quality factor is less than $%
10^{4} $. After similar analysis, we find the value of the term $\nu
_{h}/\left( 2\mu Q\right) $\ in Eq. (\ref{10}) has the order of magnitude
greater than $10^{0}$, but the order of magnitude of the other part Re$(\xi
c_{1}^{h}c_{2}^{h\ast })$\ in Eq. (\ref{10}) is less than $10^{-1}$\ due to
the atomic dephasing $\xi \leq 1$. Therefore, under current experimental
capability, the Circuit QED system can not also be used to implement the
QHE, either.

We therefore conclude, based on the present experimental accessibility, we
will have to improve the quality factor of cavity $Q$ to a much higher level
before we can implement the PCE as a practical QHE.

\section{CONCLUSIONS}

In summary, we revised the PCE proposed in Ref. \cite{Scully-science}, and
we found that the PCE efficiency decrease monotonically with $Q$\ due to\
the cavity loss. In the ideal case, $Q\rightarrow \infty $, we regained the
result of Ref. \cite{Scully-science}, where the improvement of the working
efficiency is due to the non-equilibrium state preparation. Thus the
obtained efficiency, which is beyond the classical Carnot efficiency, does
not imply the violation of the second law of thermodynamics. This
observation has been made in Refs. \cite{Kieu,Kieu1} for different physical
system. We also phenomenologically considered atomic dephasing and found
that a short time dephasing may make the PCE become a classical one. From
these heuristic discussions, we conclude that both the photon dissipation
and atomic dephasing can diminish the efficiency of the QHE. Considering the
continue improvement of the cavity quality factor $Q$, we believe the
crucial issue in future experiments will be to control the atomic coherence.
We also would like to point out that the dissipation mechanism of atoms due
to its coupling with the environment, e.g., the vacuum modes, are not
considered microscopically in this paper. Detailed investigation of the PCE
with atomic dissipation will be presented in our forthcoming paper.

We thank Y. X. Liu and T. D. Kieu for helpful suggestions. This work is
supported by the NSFC with grant Nos. 90203018, 10474104 and 60433050, and
the National Fundamental Research Program of China with Nos. 2001CB309310
and 2005CB724508.

\end{document}